\documentclass[10pt,twocolumn,aps,superscriptaddress,showpacs,showkeys]{revtex4}
\usepackage{slashed}
\usepackage{amsmath}

\begin{document}
\title{Possible Realization of non-BCS type Superconductivity in gapped Graphene}

\author{Prasanta K. Panigrahi}
\email[Email:]{prasanta@prl.res.in}
\affiliation{Physical Research Laboratory, Navrangpura, Ahmedabad - 380 015, India}
\affiliation{Indian Institute of Science Education \& Research (IISER) - Kolkata, Salt Lake, Kolkata - 700 106, India}

\author{Vivek M. Vyas}
\email[Email:]{vivek@iiserkol.ac.in}
\affiliation{Indian Institute of Science Education \& Research (IISER) - Kolkata, Salt Lake, Kolkata - 700 106, India}

\author{T. Shreecharan}
\email[Email:]{shreet@imsc.res.in}
\affiliation{The Institute of Mathematical Sciences, C.I.T. Campus, Taramani, Chennai - 600 113, India}

\begin{abstract}
We show that the gauge field induced due to non-uniform hopping, in gapped graphene, can give rise to a non-BCS type of superconductivity. Unlike the conventional mechanisms, this superconductivity phenomena does not require any pairing. We estimate the critical temperature for superconducting-to-normal transition via Berezinskii-Kosterlitz-Thouless mechanism. Possibility of observing the same in ultra cold atomic gases is also pointed out.
\end{abstract}


\maketitle

Graphene is an atom thick allotrope of carbon, which has unusual electronic properties, owing to its hexagonal honeycomb lattice structure. It was first shown by Wallace that, the valence band and conduction band touch each other at six points in the momentum space, around which quasiparticle dispersion is linear \cite{wallace}. Later, Semenoff demonstrated that, the low energy quasiparticle excitations around these points, out of which only two are inequivalent, actually satisfy the massless Dirac equation \cite{semenoff}. Here, the Lorentz symmetry is an emergent feature, due to the honeycomb structure of the material, rather than a true space-time symmetry, and the speed of light here is replaced by Fermi velocity of the electrons. Experimental isolation of this material opened a way to simulate relativistic quantum physics and check validity of various fertile ideas \cite{novoselov}. Infact, one of the very important consequences of relativistic quantum physics, Klein paradox, was experimentally seen for the first time, certifying the validity of the underlying physics \cite{katsnelson}. Amongst other interesting effects, that make this material important, are room temperature quantum Hall effect \cite{qhe},  Andreev reflection \cite{beenaker}, and universal conductance \cite{review2,rmp}. Graphene, being intrinsically a two dimensional system, can lead to realization of peculiar effects, only possible in the planar world, like fermion number fractionalization due to vortices \cite{jackiw1} and anyonic excitations \cite{jackiw2}.

Possibility of superconductivity in graphene has attracted considerable attention from both theoretical and experimental side \cite{super1,wilczek}. However, the search for superconductivity in pure graphene has yielded null results, till this date. Proximity induced superconductivity in graphene has been experimentally seen, indicating that graphene has phase coherence \cite{indsup}. The possibility of superconductivity in pure and doped graphene, and graphitic layers, has been theoretically investigated, using variants of conventational pairing mechanisms \cite{sonin,castroneto}. In Ref. \cite{baskaran}, the authors address the possibility of high $T_{c}$ superconductivity using resonanting valence bond model for doped graphene. Doped graphite is known to be superconducting, and possible mechanisms giving rise to this effect have been proposed \cite{weller,super2}. In Ref. \cite{tanaka}, the authors propose a mechanism of superconductivity arising due the edge states in graphene.

In this letter, we show that, apart from the conventional BCS type pairing mechanisms, two dimensional nature of graphene can make it superconducting via an unconventional non-BCS type mechanism. We show that, this mechanism does not require any short range four fermi interaction usually associated with pairing. Rather this mechanism crucially relies on presence of a gauge field, which may arise from lattice deformation, which locally modify the hopping amplitudes. It is shown that, existence of such a gauge field interacting with massive Dirac fermions, where mass may arise due several plausible mechanisms, via one loop correction can couple to external photon field through a topological Chern Simons coupling. This topological coupling induces a parity preserving non-topological mass for the photon field, giving rise to perfect diamagnetism or Meissner effect, which is a hallmark of superconductivity. Further, planar nature of this model ensures logarithmic confinement of all the vortex excitations present in the induced gauge field, untill a critical temperature, above which the via Berezinskii-Kosterlitz-Thouless mechanism, the vortex unbinding takes place and superconducting state is destroyed. We estimate the critical temperature for this model and show that it is proportional to the gap. We conclude by pointing out a possible way of realizing the above phenomenon in ultra cold atomic gases, in optical lattice with tunable interactions.

In tight binding approximation, the electron hopping Hamiltonian, defined on a hexagonal lattice with hopping energy $t_{i}(\bf{r})$, reads:
\begin{equation}\label{h0}
H =-\sum_{\bf{r}} \sum_{\bf{i=1,2,3}} t_{i}({\bf{r}}) \left (  a^{\dagger}({\bf{r}}) b({\bf{r}} + {\bf{s}_{i}}) + b^{\dagger}({\bf{r}}+{\bf{s}_{i}}) a({\bf{r}})  \right ), 
\end{equation}
where the fermion operators $a$ and $b$ act on the two interpenetrating sublattices A and B \cite{convention1}. The hopping energy may be written as $t_{i}({\bf{r}}) = t + \delta t_{i}(\bf{r})$, where $t$ is constant, so that the above Hamiltonian can be written conveniently as:
\begin{equation}
H = H_{D} + H_{G}
\end{equation}
where,
\begin{eqnarray}\nonumber
H_{D} =  -\sum_{\bf{r}} \sum_{\bf{i=1,2,3}} t \left (  a^{\dagger}({\bf{r}}) b({\bf{r}} + {\bf{s}_{i}}) + b^{\dagger}({\bf{r}}+{\bf{s}_{i}}) a({\bf{r}})  \right ),\\ \nonumber
H_{G} =  -\sum_{\bf{r}} \sum_{\bf{i=1,2,3}} \delta t_{i}({\bf{r}}) \left (  a^{\dagger}({\bf{r}}) b({\bf{r}} + {\bf{s}_{i}}) + b^{\dagger}({\bf{r}}+{\bf{s}_{i}}) a({\bf{r}})  \right ).
\end{eqnarray}
As is well known, that $H_{D}$ can be linearized around two Dirac points $K_{\pm}$, to yield: 
\begin{eqnarray} \nonumber
\nonumber H_{D} = -i v_{F} \int d^{2}x \left\lbrace   \Psi^{\dagger}_{+}({\bf{r}}) {\bf{\sigma}} \cdot {\nabla} \Psi_{+}({\bf{r}}) \right. \\ \nonumber \left. + \Psi^{\dagger}_{-}({\bf{r}}) {\bf{\sigma^{\ast}}} \cdot {\nabla} \Psi_{-}({\bf{r}}) \right\rbrace,  
\end{eqnarray}
where $v_{F}= \frac{3 t l}{2}$ is the Fermi velocity (which is hence forth set to unity, along with $\hbar$), $\Psi^{\dagger}_{\pm}({\bf{r}}) = (a^{\dagger}_{\pm}({\bf{r}}),b^{\dagger}_{\pm}({\bf{r}}))$ and Pauli matrices are defined as ${\bf{\sigma}}=(\sigma_{x},\sigma_{y})$ \& ${\bf{\sigma^{\ast}}}=(\sigma_{x},-\sigma_{y})$ \cite{wallace,semenoff}. Above Hamiltonian accurately describes the dynamics of electrons in an ideal flat graphene, free from any disorders or dislocations. The effects of disorder or deformation are captured by $H_{G}$, since deformations would lead to change in the hopping amplitudes locally, thereby giving rise to $\delta t_{i}(\bf{r})$. It is shown in Ref. \cite{rmp,saito}, that if $\delta t_{i}(\bf{r})$ is smooth over lattice spacing scales, then the two Dirac points are not coupled by disorder, and hence one can write :
\begin{eqnarray}\nonumber
\nonumber H_{G} = \int d^{2}x \left\lbrace \Psi^{\dagger}_{+}({\bf{r}}) {\bf{\sigma}} \cdot {\bf{a}}({\bf{r}}) \Psi_{+}({\bf{r}}) \right. \\ \nonumber \left. - \Psi^{\dagger}_{-}({\bf{r}}) {\bf{\sigma^{\ast}}} \cdot {\bf{a}} ({\bf{r}}) \Psi_{-}({\bf{r}})   \right\rbrace,  
\end{eqnarray}
where ${\bf{a}}=(a_{1},a_{2})$ is deformation induced gauge field, and is defined as
$a_{1}({\bf{r}}) = \delta t_{1}({\bf{r}}) - \frac{1}{2} ( \delta t_{2}({\bf{r}}) + \delta t_{3} ({\bf{r}}))$ and $a_{2}({\bf{r}}) = \frac{\sqrt{3}}{2} (\delta t_{2}({\bf{r}}) - \delta t_{3}({\bf{r}}) )$. It is very interesting to note, that the disorder couples to fermion field as a genuine U(1) gauge field, since $\delta t_{i}$s are defined on the links connecting the two sublattices, and unlike other cases where the effects of disorder can be captured by averaging method or replica method. The appearance of disorder as an effective gauge field, is not astonishing. It is known that effects of dislocations (like presence of an odd pentagon/heptagon in the honeycomb lattice), local curvature and phonon-electron interactions can be treated as gauge fields \cite{rmp,saito}. Interestingly, certain phonons in graphene couple to matter current as $\bar{\psi} \gamma^{\mu} A_{\mu} \psi$, rather than coupling as $\bar{\psi}\phi \psi$, giving rise to some interesting effects that were seen experimentally by Raman spectroscopy \cite{saito,pisana, yan}.

In analogy with Dirac equation for massless particles, one can unify the two spinors $\Psi_{\pm}$ by working in a reducible representation of Clifford algebra, and defining
$\psi = (b_{+},-b_{-},a_{-},a_{+})^{T}$, so that the full Hamiltonian (density) now reads: 
\begin{equation}
\mathcal{H} = \bar{\psi} \left ( - i \gamma^{i} \nabla_{i} - \tilde{g} \gamma^{3} \gamma^{5} \gamma^{j} a_{j} \right ) \psi.
\end{equation} 
where $i,j$ run from $1,2$ (sum over repeated indicies is implied here and hence forth). In above Hamiltonian, we have introduced explicitly coupling constant $g$, which specifies the strength of the fermion-gauge field interaction, and we have used the standard Dirac representation for $\gamma$ matrices as :  
\begin{eqnarray}
\gamma^{0} =  \left( \begin{array}{cc} {\bf{1}} & 0 \\ 0 & -{\bf{1}} \end{array} \right), \\
\gamma^{i} =  \left( \begin{array}{cc} 0 & \sigma_{i} \\ -\sigma_{i} & 0 \end{array} \right) \, (i=1,2,3), \\
\gamma^{5} =  \left( \begin{array}{cc} 0 & {\bf{1}} \\ {\bf{1}} & 0 \end{array} \right), 
\end{eqnarray}
and $\sigma_{i}'s$ are the Pauli matrices. Taking into account, the non-dynamical component $a_{0}$, to ensure that the ground state to start with, has zero charge density, the Lagrangian corresponding to above Hamiltonian, with a mass term, may be written in relativistic form as:
\begin{equation}\label{lag}  
\mathcal{L} = \bar{\psi} ( i \slashed{\partial} - m  + \tilde{g} \gamma^{3} \gamma^{5} \slashed{a} ) \psi,
\end{equation}
which has an emergent Lorentz symmetry. Interestingly, the matter current, $\bar{\psi} \gamma^{\mu} \gamma^{3} \gamma^{5} \psi$ coupled to the gauge field is conserved, even in the presence of a mass term. The corresponding conserved charge is $Q= \int d^{3}x  \left\lbrace  ( a^{\dagger}_{+} a_{+} + b^{\dagger}_{+} b_{+} ) - ( a^{\dagger}_{-} a_{-} + b^{\dagger}_{-} b_{-}) \right\rbrace $.

In above Lagrangian, we have introduced a mass term, which does not follow from the hopping Hamiltonian. The massless nature of the fermionic excitations in graphene in presence of a substrate have been confirmed conclusively from experiments \cite{rmp}. However, recent computational studies by Drut \& Lahde has attracted significant attention, who showed that free suspended graphene in vaccum, may well be an insulator with a gap, in a resonanting valence bond ground state \cite{drut1,drut2,drut3,netophy}. Earlier, on the grounds of symmetry and renormalization group analysis, it was shown that at low energies, graphene may break chiral symmetry and can become gapped \cite{herbut}. Unfortunately, the experimental study on freely suspended graphene sheets have been difficult and hence the nature of its ground state is hence unclear. Also, it has been shown that, graphene can infact be made gapped, that too by several ways, for example by inducing local potentials \cite{netophy}. Hence, consideration of gapped graphene spectrum is not an unphysical assumption.

The external photon field, living in $3+1$ D space time, interacting with fermions confined on a plane can not be correctly described by $-\frac{1}{4}F_{\mu \nu} F^{\mu \nu}$ ($\mu$, $\nu=0,1,2$), since it has only one degree of freedom,  and not all components of physical photon field couple to matter. Following the arguments of Ref. \cite{kovner}, we begin with photon field action $\int d^{4}x \frac{-1}{4}F_{\mu \nu} F^{\mu \nu} + j_{\mu} A^{\mu}$, where $j^{\mu}$ describes the matter current confined to XY-plane. Integrating out z-coordinate in the above action, and using Greens function identity, we get the three dimensional action as : $\int d^{3}x  \frac{-1}{4}F_{\mu \nu} \frac{1}{\sqrt{\partial^{2}}}F^{\mu \nu} + j_{\mu} A^{\mu} $, where $F_{\mu \nu}$ ($\mu$, $\nu=0,1,2$) describes 3D physical magnetic field $B_{z}$, in-plane components of electric field $E_{x}$ and $E_{y}$. Hence, the above Lagrangian suitably coupled to photon field reads:

\begin{equation} 
\mathcal{L} = \bar{\psi} ( i \slashed{\partial} - m + \tilde{g} \gamma^{3} \gamma^{5} \slashed{a} + \slashed{A}) \psi 
- \frac{1}{4 g^{2}} F_{\mu \nu} \frac{1}{\sqrt{\partial^{2}}} F^{\mu \nu}.
\end{equation}
It needs to be pointed out here that, we have assumed velocity of light $c$ equal to the Fermi velocity $v_{F}$ which is set to unity. This is actually not the case, but as shown in Ref. \cite{dorey,kovner}, this does not change the qualitative behaviour of the theory. Also, notice that the coupling $\tilde{g}$ is not dimensionless, as in $QED_{4}$, but rather has dimension of square root of mass, and the potential between two such charges is logarithmic, as expected.  
Next we integrate out the fermion fields, and find one loop contribution to effective action using derivative expansion of fermion determinant \cite{pkp1,das}, as :
\begin{equation}
{\mathcal{L}}_{eff} = - \frac{1}{24 {\pi}^{2} |m|} {f_{\mu \nu}} {f^{\mu \nu}} - \frac{ \tilde{g} m} {2 \pi |m|} \epsilon^{\mu \nu \rho} A_{\mu} {\partial}_{\nu} {a}_{\rho}.
\end{equation}
As one can clearly see, the two gauge fields are coupled by a mixed Chern-Simons term, which has a topological nature \cite{deser}. Note that the pure Cherm Simons term for both gauge fields, in this case is absent, as $Tr({\gamma^{\mu} \gamma^{\nu} \gamma^{\rho} \gamma^{3} \gamma^{5}})$ is the only nonvanishing trace of three gamma matrices in this representation, and gives rise to mixed Chern-Simons term. Also note that the sign of $m$ appears as the coefficient of this mixed term, and hence absolute magnitude of the mass is irrelevant for this mechanism, and of course massless fermions will forbid this coupling. The Chern-Simons coupling, may lead one to conclude that parity must be violated in this interaction. However, by defining $A_{\pm} = A \pm a$ fields, the Chern-Simons term in above Lagrangian can be rewritten, modulo a constant, as : $ \epsilon_{\mu \nu \rho} A^{\mu}_{+} {\partial}^{\nu} {A}^{\rho}_{+} - \epsilon_{\mu \nu \rho} A^{\mu}_{-} {\partial}^{\nu} {A}^{\rho}_{-}$, which has two topological terms with opposite signs; as a result it does not violate parity. It is straightforward to see, by integrating out $a_{\mu}$, that the above coupling induces a {\emph{parity conserving non topological}} mass $M = \tfrac{3 {\tilde{g}}^{2} |m|}{4}$ for the physical electromagnetic field and leads to spontaneous breaking of the $U_{E}(1)$ symmetry of the electric charge. The photon field becoming massive implies Meissner effect, a hallmark of superconductivity, where the static magnetic field exponentially dies down with distance from the boundary, with characteristic length scale, called penetration depth, which in our case is $\lambda = \tfrac{4}{3 {\tilde{g}}^{2} |m|}$. To make the comparison with usual BCS theory clear, we follow the technique given in Ref. \cite{banks}, and define an auxillary scalar field $\phi$, such that $\partial_{\mu} \phi = \epsilon_{\mu \nu \rho} \partial^{\nu} a^{\rho}$, so that the above Lagrangian after a Hubbard-Stratonovich transformation, reads:
\begin{equation}
{\mathcal{L}}_{eff} = {12 {\pi}^{2} |m|} \left( {\partial}_{\mu} \phi - \frac{ \tilde{g} m}{4 \pi |m|} A_{\mu}  \right)^{2}. 
\end{equation}
As we can clearly see, the above Lagrangian is in manifest London form, and all the phenomenological properties of superconductivity would follow from here \cite{wein}. It is worth pointing out, that this is a non-BCS type of superconductivity, and is present only in the planar $2+1$ D world. 

Following the arguments given in Ref. \cite{kovner,pkp2}, one can see that the correspondence between $a_{\mu}$ and $\phi$ ceases to be one-to-one when vortex excitations are present in $a_{\mu}$, and the above Lagrangian has to be corrected to consider this effect. It can be done by substituting ${\partial}_{\mu} \phi$ by ${\partial}_{\mu} \phi - i \varphi^{*} {\partial}_{\mu} \varphi$, where $\varphi^{*} \varphi =1$. 
Defining vortex current as: $J^{vort}_{\mu} = (2 \pi i)^{-1} \epsilon_{\mu \nu \lambda} \partial^{\nu} (\varphi^{*} \partial^{\lambda} \varphi)$, allows us to write the Lagrangian :
\begin{eqnarray}
\nonumber {\mathcal{L}}_{eff} = - {24 \pi^2 |m|} J^{vort,\mu} \frac{1}{\partial^2} J^{vort}_{\mu} \\ 
- 6 {\pi}^{2} m {\tilde{g}} \epsilon_{\mu \nu \lambda} J^{vort,\mu} \frac{1}{\partial^2} {\partial}^{\nu} A^{\lambda} -  \frac{3 |m| {\tilde{g}}^{2}}{16} F_{\mu \nu} \frac{1}{{\partial^{2}}} F^{\mu \nu} 
 \end{eqnarray}

Notice that the last term is topologically trivial, and is responsible for superconductivity. At low temperatures, the vortices are bound and at length scales larger than the binding scale, photon field sees vortex neutral current {\emph{i.e.,}} $J^{vort}_{\mu} \approx 0$. Hence, there is no contribution from the vortex terms in the above Lagrangian and one has a superconducting phase. The vortex bound states have finite energy, and at a finite temperature $T_{BKT}$ the vortex unbinding takes places, known as Berezinskii-Kosterlitz-Thouless phase transition in the literature \cite{ber,kt}. Considering a finite vortex current, one can show that the vortex contribution to photon field is $\frac{3 |m| {\tilde{g}}^{2}}{16} F_{\mu \nu} \frac{1}{{\partial^{2}}} F^{\mu \nu} $, which precisely cancels the contribution from non-vortex part, hence destroying superconductivity. So, the superconducting-to-normal phase transition occurs at temperature $T=T_{BKT}$, which is given by following transcendental equation:
\begin{equation}
 \mathrm{tanh} \left( \frac{|m| \beta_{BKT}}{2} \right) \beta_{BKT} = \frac{1}{6 \pi^3 |m|}.
\end{equation}
This equation, when numerically solved, gives $T_{BKT}=\tfrac{|m|}{\alpha}$, where $\alpha=0.1037312$ \cite{num}. As one can clearly see, $T_{BKT} \propto |m|$.

We have found that, disorder can couple to massive fermionic quasiparticles in graphene as a gauge field, to give rise to a special kind of superconductivity. As pointed out by Sasaki and Saito, such a gauge field can also rise from certain phonon modes \cite{saito}. In usual pairing mechanisms, which describe low temperature superconducters, an effective four fermi interaction leads to creation of fermion-fermion pairs, which successively Bose condense to give a new superconducting ground state, which has lower symmetry than the Lagrangian. Hence, in such case, superconducting state is a new vacuum, which is unitarily inequivalent to normal perturbative vacuum, which has free fermionic excitations asymptotically \cite{umezawa}. Further, the superconducting vacuum is a condensate of fermion pairs, and is a macroscopic object. Certain excitation of this condensate, so called Amplitude Mode, have been seen experimentally, establishing unambigiously its existence. In sharp contrast to this, the current proposal for superconductivity in graphene, is not described by any kind of pairing whatsoever. We show, that the perturbative ground state of gapped graphene, in presence of a gauge field which may arise from disorder, exhibits superconductivity. This directly implies absence of any amplitude mode, since there is no condensate in this model. This feature makes this type of superconductivity experimentally distinguishable from the conventional ones.

An interesting and very promising possibility of observing this type of superconductivity is in ultra cold fermionic atoms trapped in two dimensional optical lattices having hexagonal honeycomb structure. It is recently proposed that such a system would precisely follow the abovementioned analysis and would have an emergent Dirac equation governing the dynamics \cite{colddirac}. Further, the authors also show that mass of the Dirac particles is governed by the anisotropy in the hexagonal optical lattices, which can be well tuned using laser light. This opens up a possibility of tuning the superconducting transition temperature $T_{BKT}$, since it varies as $|m|$. In that case it would be interesting to see, if it is experimentally possible to create the required gauge field using appropriate light-matter interactions. This would provide a clean and decisive test for this exotic form of superconductivity. The celebrated Berezinskii-Kosterlitz-Thouless phase transition has been recently seen in ultra cold atomic system, strengthening further, the hope of realization of this form of superconductivity in ultra cold media \cite{bktcold}.  

\noindent \textit{Acknowledgements-} VMV thanks Dr. Said Sakhi for many profitable discussions. VMV also thanks Prof. Ashok Das for many enlightening discussions and for his critical comments on this work.

\end{document}